\documentclass[a4paper,12pt]{article}
\usepackage{a4wide}

\usepackage{amsmath,amssymb, amsfonts}
\usepackage{graphicx}
\makeatletter
\renewcommand\section{\@startsection {section}{1}{\z@}%
                                   {-3.5ex \@plus -1ex \@minus -.2ex}%
                                   {2.3ex \@plus.2ex}%
                                   {\large\bf}}   
\renewcommand\subsection{\@startsection {subsection}{1}{\z@}%
                                   {-3.5ex \@plus -1ex \@minus -.2ex}%
                                   {2.3ex \@plus.2ex}%
                                   {\normalfont\bf}}         
\makeatother
\allowdisplaybreaks
\makeatletter
\newcommand{\fmslash}[2][0mu]{%
  \mathchoice
    {\fmsl@sh\displaystyle{#1}{#2}}%
    {\fmsl@sh\textstyle{#1}{#2}}%
    {\fmsl@sh\scriptstyle{#1}{#2}}%
    {\fmsl@sh\scriptscriptstyle{#1}{#2}}}
\newcommand{\fmsl@sh}[3]{%
  \m@th\ooalign{$\hfil#1\mkern#2/\hfil$\crcr$#1#3$}}
\makeatother

\def\bra#1{\mathinner{\langle{#1}|}}
\def\ket#1{\mathinner{|{#1}\rangle}}
\def\braket#1{\mathinner{\langle{#1}\rangle}}
\newcommand{\sbraket}[1]{\lbrack #1\rbrack}

\DeclareMathOperator{\tr}{tr}
\newcommand{\ii}{\mathrm{i}}

\newcommand{\CP}{\mathbb{CP}}
\newcommand{\dbar}{\bar\partial}

\allowdisplaybreaks
\numberwithin{equation}{section}

\begin{document}

%%%%%%%%%%%%%%%%%%%%%%%%%%%%%%%%%%%%%%%%%%%%%%%%%%%%%%%%%%%%%%%%%%%%%%%%
\thispagestyle{empty}
\begin{flushright}
{\small
PITHA~07/23 \\
SFB/CPP-07-93\\
arXiv:0712.3409[hep-th] 
\hfill\\
December 20, 2007
}
\end{flushright}

\vspace{\baselineskip}

\begin{center}
{\LARGE CSW rules for a massive scalar}\\
 \vspace{2\baselineskip}
{\large Rutger Boels}\thanks{boels@nbi.dk}\\
{\normalsize\it Niels Bohr Institute}\\
{\normalsize\it Niels Bohr International Academy}\\
{\normalsize\it Blegdamsvej 17, DK-2100 Copenhagen, Denmark}\\
{\normalsize and}\\
{\normalsize\it The Mathematical Institute, University of Oxford}\\
{\normalsize\it 24-29 St. Giles, Oxford OX1 3LP, United Kingdom}\\
\hfill\\
\large{Christian Schwinn}\thanks{schwinn@physik.rwth-aachen.de}\\
{\normalsize \it  Institut f\"ur Theoretische Physik E}\\ 
{\normalsize \it RWTH Aachen, D - 52056 Aachen, Germany}\\\hfill\\
\end{center}

 \date{}

%\maketitle

\setcounter{page}{0}

\begin{abstract}
\noindent 
We derive the analog of the Cachazo-Svr\v{c}ek-Witten (CSW) diagrammatic Feynman rules for four dimensional Yang-Mills gauge theory coupled to a massive colored scalar.  The mass term is shown to give rise to a new tower of vertices in addition to the CSW vertices for massless scalars in non-supersymmetric theories. The rules are derived directly from an action, once through a canonical transformation within light-cone Yang-Mills and once by the construction of a twistor action. The rules are tested against known results in several examples and are used to simplify the proof of on-shell recursion relations for amplitudes with massive scalars.
\end{abstract}
%\maketitle

\newpage
%%%%%%%%%%%%%%%%%%%%%%%%%%%%%%%%%%%%%%%%%%%%%%%%

\section{Introduction}
Yang-Mills theory underlies all particle physics models including the Standard one, and the ability to make precise predictions for upcoming scattering experiments at for instance the LHC is therefore of paramount importance. Inspired by Witten's observations on twistor-space properties of Yang-Mills amplitudes~\cite{Witten:2003nn}, many new efficient methods for the calculation of these have become available in recent years. One example important for this paper are new Feynman-like rules proposed by Cachazo, Svr\v{c}ek and Witten (CSW)~\cite{Cachazo:2004kj} where off-shell continuations of maximally helicity violating (MHV) gluonic amplitudes~\cite{Parke:1986gb} are used as vertices in diagrams. This gives a dramatic reduction in the number of Feynman diagrams one has to calculate for a given process. 

Although originally only proposed for tree level applications, it was quickly realised that the CSW rules can also be applied to calculate the so-called cut-constructable pieces of one-loop amplitudes \cite{Brandhuber:2004yw}. They can also be extended in a straightforward way to those tree amplitudes for massless particles which are related by supersymmetry to glue~\cite{Georgiou:2004wu} and to single external massive Higgs or gauge bosons~\cite{Dixon:2004za,Bern:2004ba}. However from the point of view of phenomenology one would like to have rules for general propagating massive particles, and to find these it is important to know how they can be derived within field theory. 

For this the on-shell recursion relations of Britto, Cachazo, Feng and Witten~(BCFW) have been used to give indirect evidence~\cite{Britto:2005fq} and a direct proof~\cite{Risager:2005vk}. A second approach~\cite{Mansfield:2005yd} (see also~\cite{Gorsky:2005sf}) uses a canonical transformation to bring the Yang-Mills Lagrangian in light-cone gauge to a form which appears to involve only MHV vertices. This transformation was constructed explicitly in~\cite{Ettle:2006bw}, where it was verified that the first $5$ vertices indeed form off-shell MHV vertices. In a third approach initiated by Mason~\cite{Mason:2005zm}, the complete CSW rules were derived from an action written directly on twistor space by a specific gauge choice, where another gauge choice reduces the action to space-time form~\cite{Boels:2006ir,Boels:2007qn}. These developments have also allowed progress on the use of CSW-like methods in pure Yang-Mills theory on the one-loop level and led to several proposals for the construction of the rational parts of one loop amplitudes~\cite{Brandhuber:2006bf, Boels:2007gv, Brandhuber:2007vm}.

In this paper we will present CSW rules for massive colored scalars derived by both the canonical transformation as well as the twistor action method. Since amplitudes with massive scalars are directly related to those with massive quarks by supersymmetry~\cite{Schwinn:2006ca} our results are directly relevant for phenomenology. Furthermore, we expect that similar rules can be derived along these lines also for the full particle spectrum of spontaneously broken gauge theories. Finally, our results can provide insight into the calculation of the rational part of one-loop amplitudes in the CSW approach~\cite{Bern:1996je}.

The rest of this paper is organised as follows: In section~\ref{sec:rules} the CSW rules for massive scalars are
presented and some examples are worked out.  The two methods of derivation are sketched and compared in section~\ref{sec:derive}. Some further examples for the application are discussed in section~\ref{sec:applications}, including a simplification of the proof of the BCFW recursion relations for amplitudes including massive scalars~\cite{Badger:2005zh}. Technical details, a derivation of CSW rules resulting from an effective Higgs-gluon vertex and a detailed discussion of the equivalence of the light-cone and the twistor Yang-Mills will be given elsewhere~\cite{preparation}.

%%%%%%%%%%%%%%%%%%%%%%%%%%%%%%%%%%%%%%%%%%%%%%%
\section{The rules and examples}
\label{sec:rules}
\subsection{Notation}
A massive four-dimensional  scalar $\phi$ in the fundamental representation coupled to Yang-Mills theory is described by the Lagrangian
\begin{equation}
\label{eq:phi-lag}
\mathcal{L}_\phi
 = \frac{1}{4} F_{\mu \nu}F^{\mu\nu} + (D_\mu\phi)^\dagger D^\mu\phi-m^2\phi^\dagger \phi
\end{equation}
where $D_\mu=\partial_\mu-\ii g A_\mu$ and $A_\mu= T^a A_{a,\mu}$ with
$T^a$ the generators of the fundamental representation of the gauge
group. Below the two-component spinor notation will be used where to
every light-like four-momentum two spinors $\pi_p^{\alpha}=\ket{p-}$
and $\pi_p^{\dot\alpha}=\bra{p-}$ are associated that satisfy
$p^{\alpha\dot\alpha}=p_\mu \bar\sigma^{\mu \alpha \dot\alpha}=
\pi_p^{\alpha}\pi_p^{\dot\alpha} $. The dotted spinors will be
referred to as 'holomorphic' and the un-dotted ones as
'anti-holomorphic' ones, following the conventions of
~\cite{Boels:2007qn} which are opposite to the ones in
\cite{Witten:2003nn}.  Lorentz invariant spinor products are defined
by $\braket{ p q } = \braket{ p - | q +} = \pi_p^{\dot \beta}
\pi_q^{\dot{\alpha}} \varepsilon_{\dot{\alpha}\dot{\beta}} =
\pi_p^{\dot \beta} \pi_{q,\dot\beta}$ and $ [ q p ] = \braket{ q + | p
  - } = \pi_q^{\alpha}\varepsilon_{\alpha\beta} \pi_p^\beta = \pi_{q\,
  \alpha} \pi_p^\alpha$.  Light-cone components of the momenta are
defined by $ p_\pm = \frac{1}{\sqrt 2}(p_0 \mp p_3)$ and $p_{z/\bar z}
=-\frac{1}{\sqrt 2}( p_1 \mp \ii p_2)$. These conventions mainly
follow~\cite{Brandhuber:2006bf}. In terms of the light-cone components
the spinors can be taken as $\pi_p^{\dot \alpha} = 2^{1/4} \left(
  \sqrt{p_+}, p_{z}/\sqrt{p_+} \right)$ and $\pi_p^\alpha= 2^{1/4}(
\sqrt{p_+}, p_{\bar z}/\sqrt{p_+} )$.  All amplitudes and vertices in
this article are color-ordered~\cite{Mangano:1990by} using the same
conventions as in~\cite{Schwinn:2006ca}.  The sub-leading color
structures appearing for four or more particles in the fundamental
representation will not be considered in this paper.

\subsection{CSW vertices for a massive scalar}
As will be shown below, the CSW rules can be derived by a field
transformation from \ref{eq:phi-lag}. The new field variables are
$\bar B$, $B$, $\xi$ and $\bar{\xi}$ where $B$ corresponds to the
positive helicity gluons and $\bar B$ to negative helicity gluons. The
other fields are the scalar and it's complex conjugate which will be
treated as independent fields. In these variables the rules have the
following vertices:
\begin{align}
\label{eq:csw-glue}
  V_{\text{CSW}}(\bar B_1,B_2,\dots \bar B_{i},\dots, B_n)&=
\ii 2^{n/2-1}  \frac{ \braket{1i}^4}{
  \braket{12}\dots\braket{(n-1)n}\braket{n1}}\\
\label{eq:csw-2phi}
  V_{\text{CSW}}(\bar\xi_1,B_2,\dots \bar B_{i},\dots \xi_n)&=
-\ii 2^{n/2-1}  \frac{ \braket{in}^2\braket{1i}^2}{
  \braket{12}\dots\braket{(n-1)n}\braket{n1}} \\
  V_{\text{CSW}}( \bar\xi_1,B_2,\dots \xi_{i},\bar\xi_{i+1}\dots \xi_n)&=
-\ii 2^{n/2-2}  \frac{
\braket{1i}^2\braket{(i+1)n}^2}{
  \braket{12}\dots\braket{n1}}
 \left(1+\frac{ \braket{1(i+1)}\braket{in}}{\braket{1i}\braket{(i+1)n}}\right)
 \label{eq:csw-4phi}
\end{align}
and an additional tower of vertices with a pair of scalars and an arbitrary number of positive helicity gluons that is generated from the transformation of the mass term:
 \begin{equation}
\label{eq:csw-mass}
V_{\text{CSW}}(\bar\xi_1,B_2,\dots \xi_n)= -\ii  2^{n/2-1}
\frac{ m^2 \braket{1n}}{
\braket{12}\dots\braket{(n-1)n }} 
\end{equation}
The propagators are given by $\ii/(p^2-m^2)$ for the scalar fields and $\ii/(p^2)$ for the gluons. Off-shell spinors are defined as usual in the CSW rules~\cite{Cachazo:2004kj} using an arbitrary but fixed anti-holomorphic spinor $\eta^{a}$:
\begin{equation}
\label{eq:csw-continue}
k_{\dot \alpha}=k_{\dot\alpha\alpha}\eta^{\alpha}
\end{equation}
Spinors corresponding to on-shell massive scalars are defined in the same way.
External wave function normalisations are already included in the
vertices. In the light-cone gauge approach to the CSW
rules~\cite{Mansfield:2005yd} the spinors are defined in terms of the
light-cone components $(p_+,p_{z},p_{\bar z})$ also for off-shell
momenta. This corresponds to the off-shell continuation~\eqref{eq:csw-continue} with a fixed reference spinor
$\eta^\alpha\sim (0,1)^T$~\cite{Mansfield:2005yd} but the derivation
of the CSW rules can be extended to arbitrary off-shell
continuations~\cite{preparation}. Scattering amplitudes calculated
with the above rules will be independent of $\eta$, which follows from
both derivations.

The rules presented above, and in particular the vertex~\eqref{eq:csw-mass} generated by the mass term of the scalars are the main result of this paper. 
\begin{figure}[t]
  \begin{center}
  \includegraphics[width=0.6\textwidth]{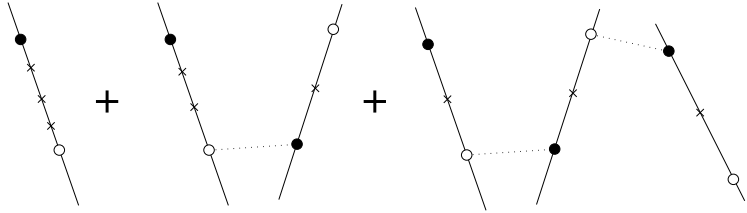}
  \caption{Twistor space structure of the amplitude with $2$ massive scalars and $3$ positive helicity gluons}
  \label{fig:twistorloc}
  \end{center}
\end{figure}
The rules differ even for \emph{massless} scalars from the supersymmetric ones considered in the literature~\cite{Georgiou:2004wu}, since in that case the space-time action contains an extra $\phi^4$ interaction. In contrast to the CSW formalism for
massless particles, the massive vertices do not correspond to off-shell continuations of on-shell scattering amplitudes. Furthermore,
in the massless scalar case the number of vertices in the CSW diagrams
is fixed to be $d\equiv n_--n_{\bar \xi}-1$ with $n_-$ the number of
external $\bar B$ lines and $n_{\bar \xi}$ that of $\bar\xi$ lines. For
massive scalars the number of massless MHV vertices~\eqref{eq:csw-glue}--\eqref{eq:csw-4phi} remains equal to
$d$ but they appear in all possible combinations with the mass-vertices~\eqref{eq:csw-mass}.

Since the vertex~\eqref{eq:csw-mass} is holomorphic, it localises on a
line ($\CP^1$) in twistor space. Therefore massive scalar amplitudes
do not localise on simple geometric structures in twistor space.
Instead, they are in general a sum of terms which localise on lines in
twistor space connected by massive propagators. This is illustrated in
figure \ref{fig:twistorloc} for an amplitude with three positive
helicity gluons. The maximum number of lines which contributes is
equal to the number of gluons in the amplitude. The failure to
localise on a simple structure in twistor space is a simple
manifestation of the fact that massive scalars are not invariant under
the conformal group.

\subsection{Examples}
\label{sec:examples}
As a check, here the rules presented above will be shown to reproduce known results for the three- and four-point amplitudes~\cite{Bern:1996ja,Badger:2005zh,Forde:2005ue}. 

The most interesting three point amplitude is that of two scalars and a positive helicity gluon. The space-time vertex which generates this is eliminated by the transformation to the new field variables but an interaction of the same field content reappears in the vertex~\eqref{eq:csw-mass} generated by the transformation of the mass
term. Since the spinors associated to the scalars are defined through~\eqref{eq:csw-continue} this vertex can be written as
\begin{equation}
\label{eq:csw-cubic}
V_{\text{CSW}}(\bar \xi_1,B_2,\xi_3)= -\sqrt 2\ii m^2
 \frac{  \braket{13}}{\braket{12}\braket{23}}
=
\frac{-\sqrt 2 \ii m^2  \braket{\eta+|\fmslash k_1\fmslash k_3|\eta-}}{
\braket{\eta+|\fmslash k_1|2+} \braket{2-|\fmslash k_3|\eta-}}   
= \frac{\sqrt 2 \ii m^2  \sbraket{2\eta}}{ \braket{2-|\fmslash k_3|\eta-}}.
\end{equation}
From the last form, it follows that this vertex vanishes if $\ket{\eta
  -}=\ket{2-}$, which will be useful in calculations below. The only
exception to this is the three-particle amplitude where the
denominator has a simultaneous pole for this choice of reference
spinor. For on-shell particles the expression~\eqref{eq:csw-cubic} is equivalent to the
vertex contained in the original action~\eqref{eq:phi-lag} written
in spinor-helicity form~\cite{Badger:2005zh}:
\begin{equation}
\label{eq:bgks}
V(\phi^\dagger_1,A_{z,2},\phi_3)= 
\sqrt 2 \ii \frac{\braket{\eta-|\fmslash k_1|2-}}{\braket{\eta 2}}
=-\sqrt 2 \ii \frac{\braket{\eta-|\fmslash k_1\fmslash k_3\fmslash k_2|\eta-}}{
\braket{\eta 2} \braket{2-|\fmslash k_3|\eta -}}
=\frac{\sqrt 2 \ii m^2\sbraket{2\eta}}{ \braket{2-|\fmslash k_3|\eta-}}
\end{equation}
 This supports a general argument \cite{preparation} that there are no equivalence theorem violations for \emph{massive} particles. 
Note that the vertex~\eqref{eq:bgks} as well as~\eqref{eq:csw-cubic}
is only independent of the choice of
$\eta$ if the external particles are on-shell~\cite{Badger:2005zh}.
It is easy to see that the other three-point vertex~\eqref{eq:csw-2phi} agrees with the result from the original action which is the conjugate of \eqref{eq:bgks}. 

For the amplitude with two positive helicity gluons and two scalars the calculation simplifies for $\ket{\eta-}=\ket{2-}$ where the second term vanishes:
\begin{equation}
\begin{aligned}
  A_4( \bar \xi_1, B_2, B_{3}, \xi_{4})
&=\frac{-2\ii m^2\braket{14}}{\braket{12}\braket{23}\braket{34}}
+\frac{-\sqrt 2\ii m^2\braket{1k_{1,2}}}{\braket{12}\braket{2k_{1,2}}}
  \frac{i}{k_{1,2}^2-m^2} 
\frac{-\sqrt 2 \ii m^2\braket{k_{1,2}4}}{
\braket{k_{1,2}3}\braket{34}}  \\
&=\frac{2\ii m^2\braket{2+|\fmslash k_{3,4}\fmslash k_{4}|2-}}{
\braket{23}\braket{3-|\fmslash k_4|2-}2(k_1\cdot k_2)}
= \frac{2\ii m^2\sbraket{23}}{\braket{23}(k_{1,2}^2-m^2)}
\end{aligned}
\end{equation}
The four-point function with one positive and one negative helicity gluon also contains a diagram with a three-gluon MHV vertex and a mass-vertex:
\begin{equation}
\begin{aligned}
\label{eq:amp-+}
  A_4( \bar \xi_1, \bar B_2, B_3, \xi_{4})
=&-2\ii \frac{\braket{12}^2\braket{24}^2}{
  \braket{12}\braket{23}\braket{34}\braket{41}}
+\frac{-\sqrt 2 \ii m^2\braket{14}}{\braket{1k_{2,3}}\braket{k_{2,3}4}}
\frac{\ii}{k_{2,3}^2}\frac{\sqrt 2 \ii\braket{ 2k_{2,3}}^3}{\braket{23}\braket{3k_{2,3}}}\\
&+\frac{\sqrt 2\ii \braket{12}\braket{2k_{1,2}}}{\braket{1k_{1,2}}}
\frac{\ii }{k_{1,2}^2-m^2} 
\frac{-\sqrt 2 \ii m^2\braket{k_{1,2}4}}{\braket{k_{1,2}3}\braket{34}}
\end{aligned}
\end{equation}
Setting $\ket{\eta-}=\ket{3-}$ only the first term survives
and the known result~\cite{Badger:2005zh,Forde:2005ue} is obtained
\begin{equation}
 A_4( \bar \xi_1, \bar B_2, B_{3}, \xi_{4})=-2\ii 
\frac{\braket{3+|\fmslash k_1|2+}}{
\braket{3-|\fmslash k_4|3-}}
\frac{\braket{2-|\fmslash k_4|3-}^2}{
  \braket{23}\braket{3+|\fmslash k_4\fmslash k_1|3-}}
=2\ii \frac{\braket{3+|\fmslash k_1|2+}^2}{
2(k_{3}\cdot k_{4})\braket{23}\sbraket{32 }}
\end{equation}

%%%%%%%%%%%%%%%%%%%%%%%%%%%%%%%%%%%%%%%%%%%%%%%%%%%%%%%%%%%%%%%%%%%%%%%%%%
\section{Derivation}
\label{sec:derive}

\subsection{Derivation from a canonical transformation}
The derivation of the CSW rules for massive scalars using a canonical transformation follows similar lines as the discussion of pure Yang-Mills theory in~\cite{Mansfield:2005yd, Ettle:2006bw}. In the light-cone gauge $A_+=0$ the Lagrangian only contains the physical components  $A_z$ (the positive helicity gluon) and $A_{\bar z}$ (the negative helicity gluon) and the scalars~\cite{Brandhuber:2006bf,preparation}.
 The only non-MHV coupling  in the gluon Lagrangian can be eliminated~\cite{Mansfield:2005yd} in favour of a tower of MHV-like couplings by transforming from the fields $A_{z}$ and the conjugate momenta $\partial_+ A_{\bar z}$ to new variables $B$ and momenta $\partial_+\bar B$. The light-cone gauge Lagrangian of the scalars~\cite{Brandhuber:2006bf,preparation} contains also a non-MHV type cubic interaction of the scalars and a positive helicity gluon that can be eliminated by an additional canonical transformation from the scalars and
the canonical momenta $\partial_+ \phi^\dagger$
to new variables $\xi$ and momenta $\partial_+\bar \xi$ together with a modification of the transformation of the conjugate gluon momentum.  
The transformation is chosen to be the same for massive and massless scalars. Using methods similar to the ones used in pure Yang-Mills in~\cite{Ettle:2006bw}, it is found to be~\cite{preparation}
\begin{equation}
\label{eq:phi-trafo}
\phi_p=\sum_{n=1}^\infty\int \prod_{i=1}^n \widetilde{d k_i}\;
 (g\sqrt 2)^{n-1}
 \frac{\braket{\eta n}B_{-k_1}\dots B_{-k_{n-1}}\xi_{-k_n}}{
\braket{\eta 1}\braket{12}\dots\braket{(n-1)n}}
\end{equation}
and an identical transformation for $\phi^\dagger$. 
 In~\eqref{eq:phi-trafo} the integration measure is defined by $\widetilde dk = dk_+ dk_z dk_{\bar z}/(2\pi)^3$ and a delta-function $(2\pi)^3\delta^3(p+\sum_ik_i)$ is kept implicit. The transformation~\eqref{eq:phi-trafo} transforms the interaction terms in the light-cone gauge Lagrangian into towers of MHV-type
vertices $\mathcal{L}^{(n)}_{\bar\xi B\dots \bar B\dots \xi}$ and
$\mathcal{L}^{(n)}_{\bar\xi B\dots\xi\bar \xi B\dots \xi}$.
Arguments of~\cite{Mansfield:2005yd} suggest that these vertices are indeed the MHV vertices for massless scalars. Since the mass term was not taken into account in the definition of the transformation~\eqref{eq:phi-trafo} it is not left invariant but is transformed into a tower of vertices with only positive helicity gluons with a vertex function $V_{\text{CSW}}$ as given in~\eqref{eq:csw-mass}.
\begin{equation}
-m^2\phi^\dagger_p\phi_{-p}= \sum_{n=2}^\infty\int \prod_{i=1}^n 
\widetilde{d k_i}g^{n-2}\;
\left(\bar\xi_{k_1}B_{k_2}\dots B_{k_{n-1}}\xi_{k_n}\right)
 V_{\text{CSW}}(\bar \xi_1,B_{2},\dots B_{n-1},\xi_n)
\end{equation}

\subsection{Twistor derivation}
In the twistor action approach~\cite{Boels:2006ir,Boels:2007qn,Boels:2007gv} off-shell gauge fields and scalars on space-time are related directly to fields on twistor space in Euclidean signature ($\CP^3 = \mathbb{R}^4 \times \CP^1$). For scalars in the fundamental representation this relation is given by
\begin{align}\label{eq:liftforphi}
\phi(x)= \int_{\CP^1} H^{-1}(\pi) \xi_0(x,\pi)\\
\phi^{\dagger}(x)= \int_{\CP^1} \bar{\xi}_0(x,\pi) H(\pi).
\end{align}
where $\pi$ parametrises the sphere. The holomorphic frame $H[B_0]$ is the solution to $(\dbar_0 + B_0) H =0$, with boundary condition $H(\eta) =0$ for some point $\eta$ on the Riemann sphere. $B_0$ and $B_{\alpha}$ are the parts of a $(0,1)$ form pointing along the sphere and space-time respectively. For more details see \cite{Boels:2007gv, preparation}. The scalar mass term is expressed in terms of the twistor-fields as
\begin{equation}
\label{eq:twistor-mass}
S_{\textrm{mass}} = - m^2 \tr \int d^4x \int_{\CP^1 \times \CP^1} \left(\bar{\xi}_0 H\right)_1\left( H^{-1} \xi_0 \right)_2
\end{equation}
 The remaining terms of the action are given by the truncation of the $\mathcal{N}=4$ twistor action \cite{Boels:2006ir} to just glue and one scalar and it's complex conjugate, evaluated in the fundamental representation. In addition one has to subtract (the lift of a) $\phi^4$ vertex contained in the $\mathcal{N}=4$ twistor action. Choosing the axial 'CSW gauge' condition~\cite{Boels:2007qn}
\begin{equation}
\eta^{\alpha} \left(B_\alpha, \xi_\alpha, \bar{\xi}_\alpha, \bar{B}_\alpha \right) = 0
\end{equation}
eliminates the interaction vertex in the holomorphic Chern-Simons term in the $\mathcal{N}=4$ action. In addition, one obtains propagators
\begin{align}
:B_0 \bar{B}_0: = \frac{\delta(\eta \pi_1 p) \delta(\eta \pi_2 p)}{p^2} \quad
:\xi_0 \bar{\xi}_0: = \frac{\delta(\eta \pi_1 p) \delta(\eta \pi_2 p)}{p^2-m^2}
\end{align}
The frame-fields can be expanded using the relation
\begin{equation}
\frac{H(\eta)H^{-1}(\pi)}{\braket{\eta \pi}} = \left(\dbar_0 + B_0 \right)^{-1}_{\eta \pi}
\end{equation} 
Using the delta-functions in the propagators the mass term~\eqref{eq:twistor-mass} is seen to lead to the vertex~\eqref{eq:csw-mass}. 
The remaining interaction vertices  in the truncated $\mathcal{N}=4$ action
 give rise to the usual MHV vertices~\cite{Boels:2007qn}. Equation \eqref{eq:phi-trafo} and \eqref{eq:liftforphi} can be seen to be equivalent when expanding out the latter and performing all sphere integrals using the delta functions. This will be discussed further in \cite{preparation}.

%%%%%%%%%%%%%%%%%%%%%%%%%%%%%%%%%%%%%%%%%%%%%%%%%%%%%%%%%%%%
\section{Simple applications}
\label{sec:applications}
As applications of the CSW representation for massive particles we consider three examples: 
a simplification of the proof of the BCFW recursion relations for massive scalars, the structure of the amplitudes with only positive helicity
gluons and a simple way to obtain the leading contribution of scattering amplitudes in the limit of a small mass.

\subsection{BCFW recursion for massive scalars revisited}
In the BCFW relations one picks two particles with momenta $k_i$ and $k_j$ and shifts the associated spinors into the complex plane. If both particles are 
massless, the shift is defined as
\begin{equation}
\label{eq:ij-shift}
\begin{aligned}
\ket{i'+}&=\ket{i+}+z\ket{j+} & 
\ket{j'-}&=\ket{j-}-z\ket{i-}
\end{aligned}
\end{equation}
If particle $j$ is massive,  
it's momentum can be decomposed into a sum of
two light-like vectors according to
$k_j =k_j^\flat +m^2/(2k_i\cdot k_j)k_i $. 
In this case the shift is defined
 as~\cite{Schwinn:2007ee}
\begin{equation}
  \ket{i'+}=\ket{i+}+z\ket{j^\flat+}\quad,\quad
 {k^\mu_j}'=k_j^\mu-\frac{z}{2}\braket{i +|\gamma^\mu| j^\flat +}
\end{equation}

The linchpin of the proof presented in~\cite{Britto:2005fq} is that the scattering amplitude
considered as a function of the complex variable $z$ must vanish as
$z\to\infty$.  For a shift of a negative and a positive helicity gluon
$(g_i^+,g_j^-)$ this can be demonstrated using an analysis of Feynman
diagrams~\cite{Britto:2005fq} while more involved methods have to be
used for shifts of particles with the same helicity~\cite{Badger:2005zh,Schwinn:2007ee}.  For pure Yang-Mills
amplitudes the validity of these shifts also follows from the CSW representation~\cite{Britto:2005fq}.

The CSW representation introduced in section~\ref{sec:rules}
allows to apply the arguments of~\cite{Britto:2005fq} to amplitudes
with massive scalars, leading to a  more direct proof of the
$(g_i^+,g_j^+)$ and $(g_i^+,\phi_j)$ shifts than in~\cite{Badger:2005zh,Schwinn:2007ee}.
 Consider the case that both gluons $(g_i^+,g_j^+)$ are connected to
 the same vertex in a CSW diagram. From the explicit form of the
 vertices~\eqref{eq:csw-glue}-\eqref{eq:csw-mass} it is easy to see
 that the diagram falls off at least as $z^{-1}$. To discuss the
 diagrams where the gluons are connected to different CSW-vertices it
 is convenient~\cite{Britto:2005fq} to use the off-shell
 continuation $\ket{\eta-}=\ket{i-}$. For this choice the spinor
 products involving internal momenta are all independent of $z$. Since
 the $z$-dependent propagators behave like $z^{-1}$ at large $z$, the
 vertex containing gluon $i$ vanishes as $z^{-2}$ and the vertex
 containing gluon $j$ is independent of $z$, it is clear that the
 amplitude vanishes as $z\to\infty$ as was to be shown.  For the shift
 of a massive scalar and a positive helicity gluon $(g_i^+,\phi_j)$
 the same off-shell continuation ensures that the CSW vertices are not
 affected by the shift of the massive scalar. The vertices containing
  gluon $i$ vanish at least as $z^{-1}$ and the same
 argument as above applies.

\subsection{Amplitudes with positive helicity gluons}
\label{sec:all-plus}

The simplest amplitudes with a pair of massive scalars are those with only positive helicity gluons.  Using the shift~\eqref{eq:ij-shift}  for $(g_2^+,g_3^+)$, they satisfy the recursion relation~\cite{Forde:2005ue,Ferrario:2006np}
\begin{equation}
 \label{eq:bcfw++} 
    A_n(\bar \xi_1,B_2,\dots,\xi_n)= A(\bar \xi_{1}, {B'}_2,
    \xi'_{K_{1,2}}) \frac{\ii}{k_{1,2}^2-m^2} A( \xi'_{K_{1,2}},
    {B'}_{3},\dots,\xi_n)
\end{equation}
 with the intermediate momentum ${K'}^\mu_{1,2}=k^\mu_{1,2}+\frac{z}{2}\braket{2+|\gamma^\mu|3+}$.
In~\eqref{eq:bcfw++} the variable $z$ is fixed to the value
$z_{1,2}\equiv -(k_{1,2}^2-m^2)/\braket{2+|\fmslash k_{1}|3+}$
in order to put $K'_{1,2}$ on-shell.
A compact solution of~\eqref{eq:bcfw++} for arbitrary $n$ has been found
in~\cite{Ferrario:2006np}.

\begin{figure}[t]
  \begin{center}
  \includegraphics[width=0.5\textwidth]{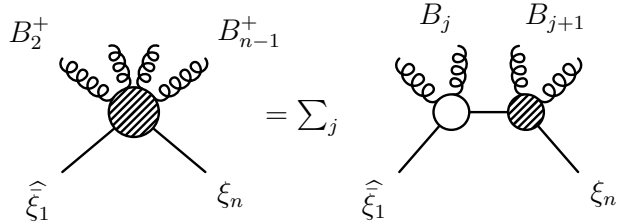}
  \caption{Recursive construction of amplitudes with
   only positive helicity gluons}
  \label{fig:csw-recursion}
  \end{center}
\end{figure}

In the CSW formalism for massive scalars, the all-plus amplitudes with a pair
of massive scalars are given by diagrams that only contain the mass-vertices~\eqref{eq:csw-mass}.
As sketched in figure~\ref{fig:csw-recursion}, 
they can be obtained recursively from a relation involving 
currents with one off-shell scalar (denoted by a hat)
\begin{equation}
\label{eq:csw-recursion}
   A_n( \widehat{\bar \xi_1},..., ,\xi_{n})
= \sum\limits_{j=2}^{n-1} 
V_{j+1,\text{CSW}}(\widehat{\bar \xi_1},B_2...,B_{j}, 
\widehat{\xi}_{-k_{1,j}})\,
 \frac{\ii}{k_{1,j}^2-m^2} 
 A_{n-j+1}(\widehat{\bar \xi_{k_{1,j}}},B_{j+1},\dots \xi_{n}) 
\end{equation}
Here the two-point function is defined as
$ A_2(\widehat{\bar \xi_{-p}},\xi_p) 
 =(-\ii) (p^2-m^2)$.
Using~\eqref{eq:csw-recursion} iteratively, the $n$-particle amplitude is expressed as a sum of diagrams with $1,2,\dots n-2$ mass-vertices, summed over all possible distributions of the gluons. This corresponds to the obvious generalisation of the twistor-space structure sketched in figure~\ref{fig:twistorloc}.

 To check that the on-shell amplitude obtained from~\eqref{eq:csw-recursion}
satisfies the relation~\eqref{eq:bcfw++}, consider a complex continuation of the amplitude, $A_n(z)$, defined  
by performing the shift for arbitrary values of the complex
parameter $z$.  For the choice $\ket{\eta-}=\ket{2-}$ for the off-shell continuation of the CSW vertex the $j=2$
term vanishes because it includes a three-point mass-vertex with the gluon $B_2$. 
In addition the $z$-dependence drops out of the spinor products involving $\ket{K_{1,2}'+}$.
 In all terms in~\eqref{eq:csw-recursion} with $j\neq 2$, the $z$-dependence comes only from the denominator of
the CSW vertex through the spinor product 
\begin{equation}
\braket{12'}\to \braket{2+|\fmslash k_1|2+}
+z\braket{2+|\fmslash k_1|3+}
={K'}^2_{1,2}(z)-m^2
\end{equation}
so it is seen that the only  pole of $A_n(z)$ is at $z_{1,2}$.
The CSW vertices  evaluated with the shifted spinors 
 factorise into the product of a CSW vertex with one leg removed, a scalar propagator
with the shifted momentum  and a three point vertex~\eqref{eq:bgks} (with $\ket{\eta+}=\ket{3+}$): 
\begin{multline}
\label{eq:csw-factorize}
V_{j+1,\text{CSW}}(\bar \xi_1,{B'}_2, {B'}_3...,B_{j}, 
\widehat{\xi}_{-k_{1,j}})
=\sqrt 2 \frac{\braket{1k_{1,j}}\braket{K'_{1,2}3}}{
\braket{12'}\braket{23}\braket{K_{1,2}'k_{1,j}}}
 V_{j,\text{CSW}}(\bar \xi_{K'_{1,2}},{B'}_3,...,B_{j}, 
\widehat{\xi}_{-k_{1,j}})\\
=\left(\ii \sqrt 2 \frac{\braket{2+|\fmslash k_1|3+}}{\braket{32}}\right)
\frac{\ii }{{K'}^2_{1,2}(z)-m^2} 
 V_{j,\text{CSW}}(\bar \xi_{K'_{1,2}},{B'}_3,...,B_{j}, 
\widehat{\xi}_{-k_{1,j}})
\end{multline}
where the CSW vertex is independent of $z$.
 Inserting this result into the 
CSW representation~\eqref{eq:csw-recursion} the sum over $j$ can be performed
to obtain the on-shell amplitude with one leg removed.
 Setting $z=0$ we obtain the
recursion relation~\eqref{eq:bcfw++} as was to be shown.

\subsection{Limit of small masses}
Since the vertex~\eqref{eq:csw-mass} is proportional to $m^2$, the
rules presented in section~\ref{sec:rules} allow to obtain the leading
piece of the amplitudes in an expansion in powers of the mass in a
simple way.  For instance, the leading contribution to the all-plus
amplitudes~\eqref{eq:csw-recursion} arises from the $n-1$ term which
contains a single vertex.  At leading order in $m^2$ the vertex is
independent of the reference spinor $\eta$.  This can be seen by
decomposing a massive momentum as $k=k^\flat +m^2/(2\eta k) \eta$ with
the same $\eta^\alpha$ as in the off-shell
continuation~\eqref{eq:csw-continue}. One can then approximate
$\fmslash k_1\ket{\eta-}\sbraket{1^\flat p_1}=\fmslash k_1\ket{p_1-}
\sbraket{1^\flat \eta}+\mathcal{O}(m^2)$ where $\ket{p_1-}$ is an
arbitrary spinor.  Using this identity for $\ket{p_1-}=\ket{2-}$ and
an analogous one for leg $n$ with $\ket{p_n-}=\ket{(n-1)-}$, 
 the spinor products in~\eqref{eq:csw-mass} that involve the
massive legs can be approximated as
\begin{equation}
  \frac{\braket{1n}}{\braket{12}\braket{(n-1)n}}
=\frac{\braket{2+|\fmslash k_1\fmslash k_{n}|(n-1)-}}{
\braket{2+|\fmslash k_1|2+}\braket{(n-1)-|\fmslash k_n|(n-1)-}}
 +\mathcal{O}(m^2)
\end{equation}
Different choices of the arbitrary spinors $\ket{p_{1/n}-}$ are equivalent at $\mathcal{O}(m^2)$.
In this way, the leading term of the all-plus
amplitudes~\cite{Bern:1996ja} is obtained from a single CSW vertex:
\begin{equation}
\label{eq:all-plus-leading}
    A_n( \bar \xi_1,B_2 ..., ,\xi_{n})=\ii 2^{n/2-1}
   \frac{- m^2 \braket{2+|\fmslash k_1\fmslash k_{n}|(n-1)- }}{
2(k_1\cdot k_2) 2(k_{n-1}\cdot k_n)
\braket{23}\dots \braket{(n-2)(n-1)}}+\mathcal{O}(m^2)
\end{equation}
  The leading piece of the amplitudes with one negative
helicity gluon is obtained from the vertex~\eqref{eq:csw-2phi}, 
in  agreement with~\cite{Forde:2005ue} up to terms of order $m^2$:
\begin{multline}
\label{eq:one-minus-leading}
    A_n( \bar \xi_1 ..., \bar B_i,... ,\xi_{n})\\=
\frac{\ii 2^{n/2-1}\braket{(n-1)+|\fmslash k_n|i+ }^2 \braket{2+|\fmslash k_1|i+ }^2  }{2(k_1\cdot k_2) 2(k_{n-1}\cdot k_n)
 \braket{2+|\fmslash k_{1}\fmslash k_n|(n-1)-}\braket{23}\dots \braket{(n-2)(n-1)}}+\mathcal{O}(m^2)
\end{multline}

%%%%%%%%%%%%%%%%%%%%%%%%%%%%%%%%%%%%%%%%%%%%%%%%%%%%%
\section{Conclusions and outlook}

In this paper we have shown that Lagrangian methods for the derivation of the CSW
rules~\cite{Mansfield:2005yd,Ettle:2006bw,Boels:2007qn} can be used to
obtain new diagrammatic rules for massive particles. As a by-product
we have also elucidated the twistor structure of massive amplitudes
and generated the complete CSW rules for general massless scalars, slightly improving on results in the literature obtained using supersymmetry~\cite{Georgiou:2004wu}.

As a first example a massive colored scalar was discussed but we expect that our methods can be extended to general spontaneously broken gauge theories. The construction differs from the MHV rules for massless particles since the vertices are not given by an off-shell continuation of on-shell amplitudes. Therefore it appears difficult to obtain our rules using the method of~\cite{Risager:2005vk} that involves only on-shell amplitudes. As an example for the usefulness of the CSW representation, it was shown to simplify the proof of the BCFW recursion relations. It was also shown how to obtain the leading piece of scattering amplitudes in an expansion in  the mass. As another application of the approach presented here, we will discuss the derivation of the CSW rules for an effective Higgs-gluon coupling~\cite{Dixon:2004za} in a forthcoming publication~\cite{preparation}.

Although the number of contributing diagrams in the massive CSW formalism is not as small as in the massless case, we believe our formalism is a significant improvement compared to the usual Feynman diagrammatic approach since all simplifications related to the purely gluonic pieces of scattering amplitudes are incorporated automatically in the vertices. For the application to large multiplicity scattering amplitudes it would be helpful to sum up the vertices with only positive helicity gluons by solving~\eqref{eq:csw-recursion}. We expect that this equation can be solved with methods similar to the ones used in~\cite{Ferrario:2006np}. Finally, through the supersymmetric decomposition, amplitudes with massive scalar-loops calculate the rational parts of one-loop Yang-Mills amplitudes and we expect that our results also provide insight into this problem. Work in this direction is in progress.

\section*{Acknowledgements}
It is a pleasure to thank Kasper Risager, David Skinner, Stefan Weinzierl and Costas Zoubos for discussions. 
We also would like to thank the Arnold Sommerfeld Center for Theoretical Physics in Munich for organising the workshop on ``Twistors, perturbative gauge theories, supergravity and superstrings'' where this work was initiated. The work of CS was supported by the DFG Sonder\-forschungs\-bereich/Trans\-regio~9 ``Computergest\"utzte Theoretische Teilchenphysik'' and by the BMBF grant 05HT6PAA. The work of RB was partly supported by the European Community through the FP6 Marie Curie RTN {\it ENIGMA} (contract number MRTN-CT-2004-5652).
%%%%%%%%%%%%%%%%%%%%%%%%%%%%%%%%%%%%%%%%%%%%%%%%%%%%%%%%%

\providecommand{\href}[2]{#2}\begingroup\raggedright
\endgroup

%%%%%%%%%%%%%%%%%%%%%%%%%%%%%%%%%%%%%%%%%%%%%%%%%%%%%%%%%%%%%%%%%%%%%%%%%%%%%

\end{document}